\documentclass{ieeeaccess}
\usepackage{cite}
\usepackage{amsmath,amssymb,amsfonts}
\usepackage{algorithmic}
\usepackage{graphicx}
\usepackage{textcomp}
\usepackage{multirow}
\usepackage{hyperref}
\def\BibTeX{{\rm B\kern-.05em{\sc i\kern-.025em b}\kern-.08em
    T\kern-.1667em\lower.7ex\hbox{E}\kern-.125emX}}
\begin{document}
\history{Date of publication xxxx 00, 0000, date of current version xxxx 00, 0000.}
\doi{10.1109/ACCESS.2017.DOI}

\title{Learning~to~Maximize~Speech~Quality Directly Using~MOS~Prediction for Neural~Text-to-Speech}
\author{\uppercase{Yeunju Choi\authorrefmark{1}, Youngmoon Jung\authorrefmark{1}, Youngjoo Suh\authorrefmark{2}, and Hoirin Kim}\authorrefmark{1},
\IEEEmembership{Member, IEEE}}
\address[1]{School of Electrical Engineering, KAIST, Daejeon 34141, South Korea}
\address[2]{Voice Group, Konan Technology Inc., Seoul 06627, South Korea}
\tfootnote{This work was supported by the National Research Foundation of Korea(NRF) grant funded by the Korea government(MSIT) (No. 2021R1A2C1014044).}

\markboth
{Author \headeretal: Preparation of Papers for IEEE TRANSACTIONS and JOURNALS}
{Author \headeretal: Preparation of Papers for IEEE TRANSACTIONS and JOURNALS}

\corresp{Corresponding author: Yeunju Choi (e-mail: wkadldppdy@kaist.ac.kr).}

\begin{abstract}
Although recent neural text-to-speech (TTS) systems have achieved high-quality speech synthesis, there are cases where a TTS system generates low-quality speech, mainly caused by limited training data or information loss during knowledge distillation.
Therefore, we propose a novel method to improve speech quality by training a TTS model under the supervision of perceptual loss, which measures the distance between the maximum possible speech quality score and the predicted one. 
We first pre-train a mean opinion score (MOS) prediction model and then train a TTS model to maximize the MOS of synthesized speech using the pre-trained MOS prediction model.
The proposed method can be applied independently regardless of the TTS model architecture or the cause of speech quality degradation and efficiently without increasing the inference time or model complexity.
The evaluation results for the MOS and phone error rate demonstrate that our proposed approach improves previous models in terms of both naturalness and intelligibility.
\end{abstract}


\titlepgskip=-15pt

\maketitle

\section{Introduction}
\label{sec:introduction}
\PARstart{S}{tate-of-the-art} text-to-speech (TTS) systems can synthesize speech that is almost indistinguishable from human speech \cite{WaveNet, ParallelWaveNet, Tacotron2, TransformerTTS}.
Nevertheless, several factors can degrade speech quality.
First, it is well known that limited training data results in the quality degradation of the synthesized speech \cite{data_reduction}. 
Therefore, system developers have needed large-scale training data to synthesize high-quality speech, despite the high cost of data collection.
Second, oversimplified or inaccurate target data during knowledge distillation can degrade speech quality.
Knowledge distillation has been proposed for TTS to improve inference speed or reduce model size \cite{FastSpeech, kd1, kd2}. 
However, some oversimplified or inaccurate data generated by a teacher model causes information loss in the target data for the student model, thus degrading the speech quality.

In this paper, we propose a novel method called perceptually guided TTS to improve the speech quality of TTS models directly.
We incorporate perceptual loss, which measures the distance between the maximum possible speech quality score and the predicted score, into the conventional training loss function for TTS. 
Here, we utilize the mean opinion score (MOS) as the quality score since it is the most widely used subjective evaluation metric for TTS. 
To predict the MOS of synthesized speech automatically, we pre-train a deep-learning-based MOS prediction model.
The proposed method is independent of the TTS model architecture or cause of speech quality degradation.
It is also efficient in that it does not increase the inference time or complexity of the model. 

\section{Related work}
\label{sec:related_work}

Many studies have proposed perceptual loss to improve the quality of outputs generated by a deep generative model.
There are generally two orthogonal approaches for defining perceptual loss. The first approach is based on style reconstruction loss proposed by Gatys \emph{et al.} \cite{Gatys}. 
It assumes that a neural network trained for classification has the perceptual information that a generative model needs to learn. 
Then, it tries to make the feature representations of the generative model similar to those of the pre-trained classification model.
Here, perceptual loss is defined as the distance between the feature representations from the generative model and those from the classification model. 
This approach has been successfully applied in various fields, including image style transfer \cite{Gatys, Johnson}, audio inpainting \cite{AudioInpaint}, speech enhancement \cite{HiFiGAN}, neural vocoding \cite{ParallelWaveNet, MelGAN}, and expressive TTS \cite{expTTS}.

The second approach uses a perceptual evaluation metric, such as the image aesthetic score, perceptual evaluation of speech quality (PESQ) \cite{PESQ}, or short-time objective intelligibility (STOI) \cite{STOI}, to learn perceptual information more directly. 
For the image enhancement task, Talebi and Milanfar \cite{ImagEn} have proposed to maximize the aesthetic score of an image enhanced by a convolutional neural network (CNN). They calculated perceptual loss using a pre-trained image assessment model and used the perceptual loss to train an image enhancement model. 
For the speech enhancement task, Zhao \emph{et al.} \cite{STOIloss} and Fu \emph{et al.} \cite{PESQloss1} have proposed to fine-tune a pre-trained speech enhancement model by maximizing the modified STOI and approximated PESQ function, respectively. 
Martín-Doñas \emph{et al.} \cite{PESQloss2} have proposed a perceptual metric for speech quality evaluation (PMSQE) by introducing two disturbance terms inspired by the PESQ algorithm. Kolbæk \emph{et al.} \cite{SElosses} investigated six loss functions including the standard loss function (i.e., mean square error) and perceptual loss function (i.e., PMSQE). Zhang \emph{et al.} \cite{SpeechSep} have proposed an approximate gradient descent algorithm to optimize PESQ and STOI directly for speech separation. 
For unit selection TTS, Peng \emph{et al.} \cite{unitselection} optimized the concatenative cost function concerning its correlation with the MOS.
For neural TTS, Baby \emph{et al.} \cite{Baby} have proposed a TTS model selection method using the phone error rate (PER) as a perceptual metric. 
In this paper, we define perceptual loss using a MOS prediction model and make a neural TTS model learn to maximize the MOS of speech.
Since we use the perceptual evaluation metric, MOS, to learn perceptual information, our method follows this second approach.

Despite a large number of prior studies using perceptual loss, only a few of those have been proposed for neural TTS. 
One of them is the work done by Baby \emph{et al.} \cite{Baby}, which selected a TTS model with the lowest PER. The authors used the PER as a criterion for selecting the best model after training is completed, not as a loss function for training the model. Our method differs from their work in two aspects: 1) we use a MOS, not the PER, as a perceptual metric, and 2) we use the perceptual metric during training, not after training. 
We argue that these two aspects make our method more advantageous for TTS.
This is because the MOS is a more widely used metric for synthetic speech evaluation than the PER, and the use of a perceptual metric during training allows a direct update of model parameters.
To the best of our knowledge, this is the first study to use MOS-based perceptual loss for training a neural TTS model.

\section{Method}
\label{sec:method}

Recent neural TTS systems generally consist of two modules: a text-to-Mel-spectrogram conversion model and a vocoder.
In this paper, we call the text-to-Mel-spectrogram conversion model the "TTS model'' for simplicity of notation and focus on the TTS model, not the vocoder.
Our method can be applied to an arbitrary TTS model regardless of the model architecture or training method since it only requires the Mel-spectrogram generated by a TTS model during training.

\begin{figure}
\centerline{\includegraphics[trim=0.2cm 0.0cm 0.3cm 0.0cm, clip=true,width=\columnwidth]{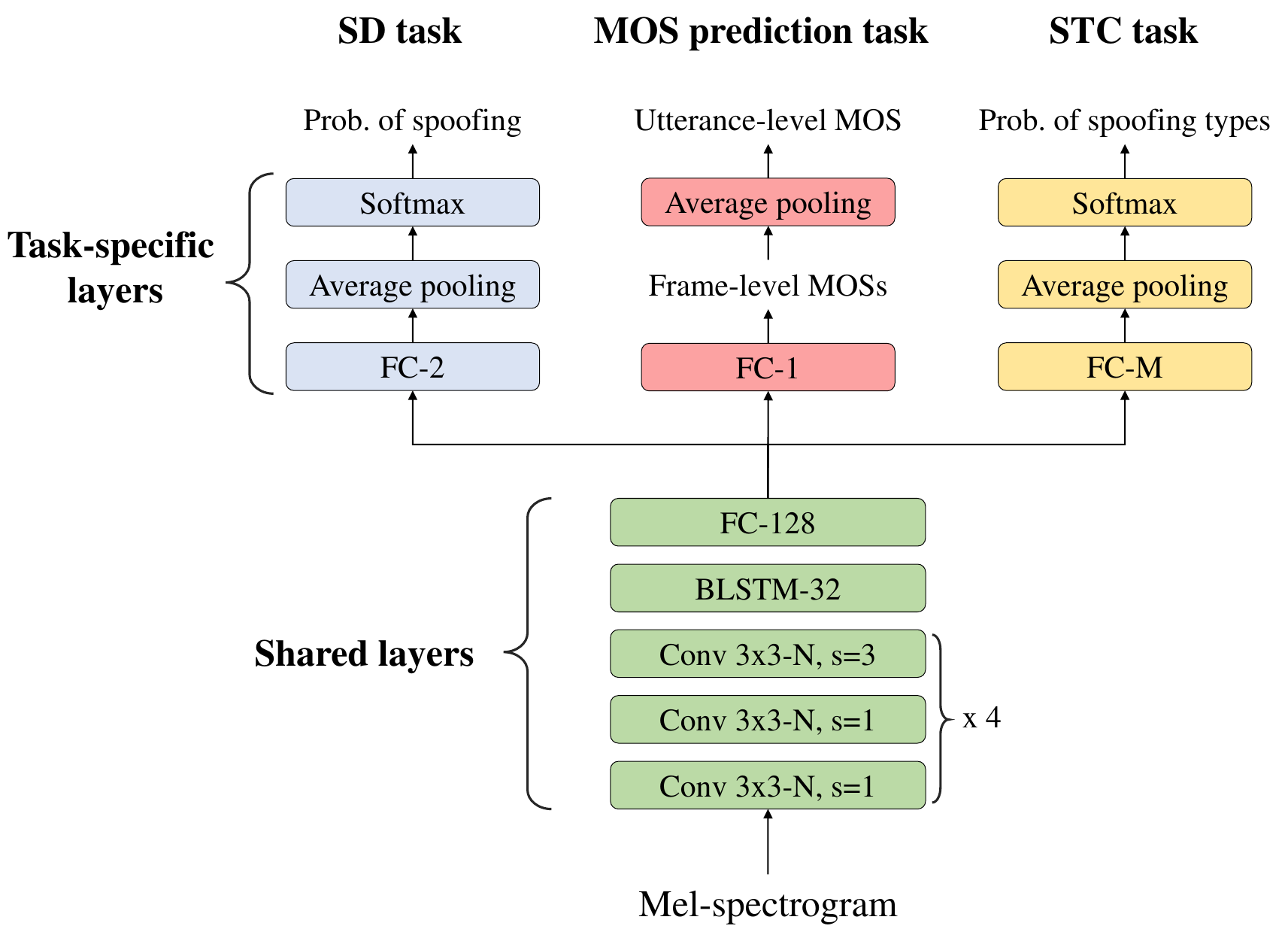}}
\caption{Overview of the MOS prediction model, MTL-MOSNet. N is the number of channels for three convolutional layers, corresponding to 16, 16, 32, and 32 for four stacks. M is the number of spoofing types and s is the stride for the convolutional layer.}
\label{fig:MOSNet_MTL}
\end{figure}

\subsection{MOS prediction model}
\label{sec:mos_prediction}
To directly improve the perceptual quality of synthesized speech, we introduce perceptual loss based on the predicted MOS.
We slightly modify MOSNet+STC+SD \cite{Choi}, an improved version of MOSNet \cite{MOSNet}, and pre-train it to predict the MOS of synthetic speech from its Mel-spectrogram. 
MOSNet is a deep neural network that predicts a MOS from a 257-dim linear spectrogram.
It consists of 12 convolutional layers, one bidirectional long short-term memory (BLSTM) layer, two fully connected (FC) layers, and a global average pooling layer.
The CNN-BLSTM network consisting of convolutional layers and a BLSTM layer acts as a feature extractor to extract frame-level features.
The outputs from the two FC layers are frame-level scores, and the final utterance-level score is obtained by the global average pooling layer.
Here, the ground truth frame-level MOSs are assumed to be the same as the ground truth utterance-level MOS and the loss function is defined as a weighted sum of frame-level and utterance-level mean square error (MSE) losses.

Fig. \ref{fig:MOSNet_MTL} shows the architecture of the MOS prediction model used in this study.
In \cite{Choi}, we proposed to use multi-task learning (MTL) with spoofing type classification (STC) and spoofing detection (SD) to improve the generalization ability of MOSNet and called the proposed model MOSNet+STC+SD. 
To begin with, we point out that we modify the model in this paper to combine it with a TTS model.
To use an 80-dim Mel-spectrogram generated by a TTS model instead of the 257-dim linear spectrogram as an input, we change the number of BLSTM units from 128 to 32.
Since synthetic speech can threaten an automatic speaker recognition system \cite{SD}, we refer to synthetic speech as ``spoofing speech'' and define a binary classification task to discriminate between human and synthetic speech as spoofing detection (SD).
STC is a multi-classification task on the subject that generates the input spectrogram, called ``spoofing type.'' 
Here, a spoofing type can be a speech generation system such as Transformer TTS and FastSpeech or a human speaker. 
Both auxiliary tasks share the feature extractor with MOS prediction except for the final FC layer (indicated by ``Shared layers'' with green color in the figure). 
Each auxiliary task has a task-specific layer that consists of an FC layer, a global average pooling layer, and a softmax layer. The task-specific layer of SD (marked by blue rectangles) outputs the probabilities of synthetic and human speech for the input spectrogram. 
The task-specific layer of STC (marked by yellow rectangles) outputs the probabilities of all spoofing types in the training data.
The total loss function for training the model is composed of four terms: both MSE losses for utterance- and frame-level MOSs and both cross-entropy losses for STC and SD tasks. 
For simplicity of notation, we denote MOSNet+STC+SD as MTL-MOSNet.

\begin{figure}
\centerline{\includegraphics[trim=1.12cm 23.3cm 1.7cm 0.3cm, clip=true,width=\columnwidth]{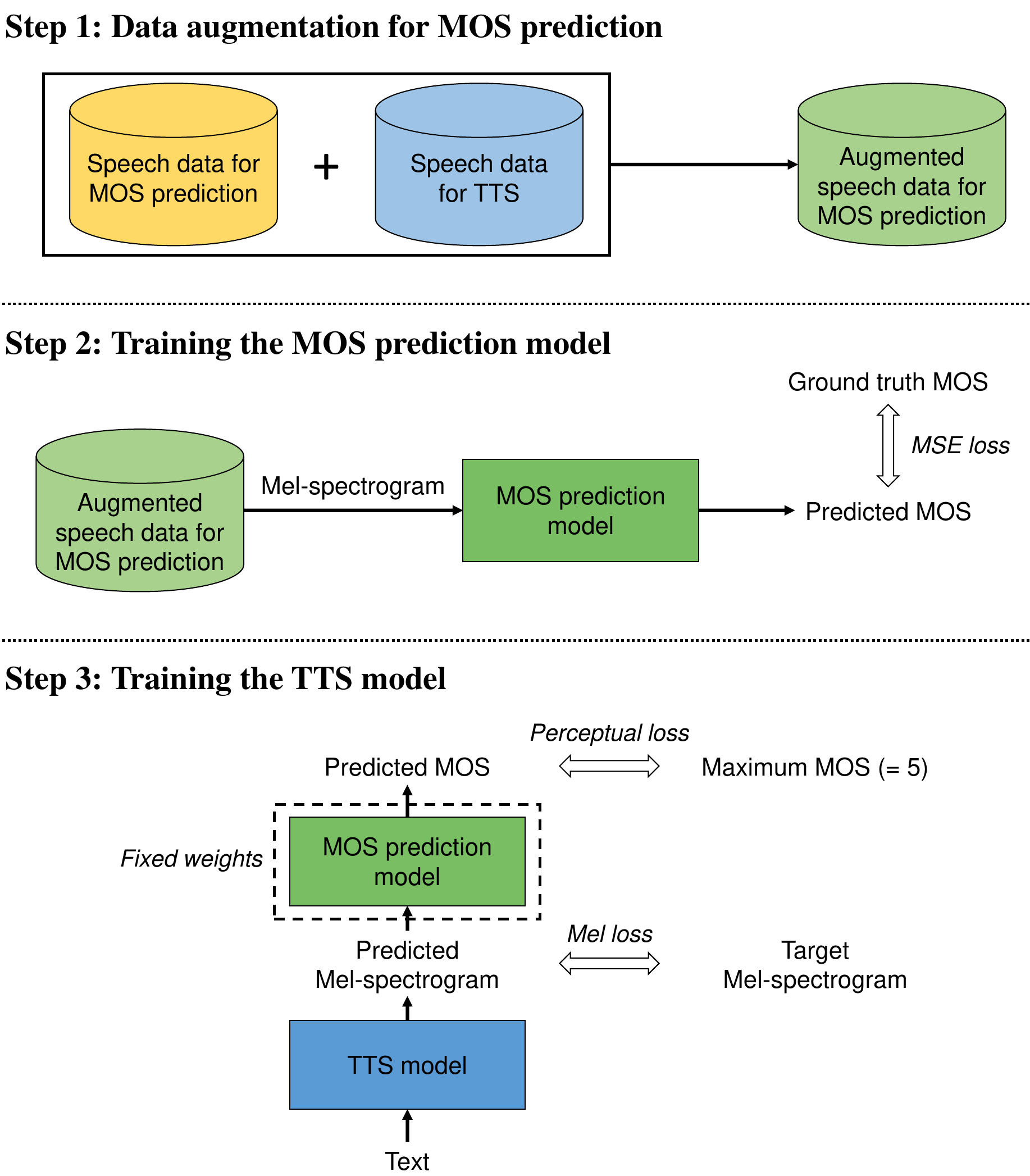}}
\caption{Overview of the proposed perceptually guided TTS with MOS prediction. In Step 3, only Mel loss and perceptual loss are shown for brevity.}
\label{fig:overview}
\end{figure}

Since there is a domain mismatch between the MOS prediction dataset and the TTS dataset, we augment the MOS prediction dataset with audio samples in the TTS dataset. In this paper, although the ``domain'' also includes the speaker and recording environment, it primarily refers to the language since we use an English dataset for MOS prediction and a Korean dataset for TTS.
Therefore, we augment with speech data uttered in the same language as the TTS dataset. 
For convenience, we only use the already-existent TTS dataset, although it is possible to augment with speech data uttered by speakers other than the target speaker.
This data augmentation process is the first step of our method, as shown in Fig. \ref{fig:overview}.
The next step is to train the MOS prediction model to minimize the MSE loss between the ground truth MOS and the predicted MOS on the augmented training data.
Here, we assume that all audio samples in the TTS dataset have a ground truth MOS of 5 since obtaining exact ground truth MOSs by a subjective test is expensive and time-consuming.
This assumption is reasonable in this paper because the TTS dataset we use was recorded by a professional speaker in a clean environment. 
Nevertheless, if the speech data in a TTS dataset was not recorded in a perfect environment, it might be better to perform a small MOS test and assume that the average of the resulting scores is the ground truth MOS.

\subsection{Perceptually guided TTS}
\label{sec:perceptually_guided_TTS}
After pre-training the MOS prediction model, we use it to calculate the perceptual loss for TTS (see Step 3 in Fig. \ref{fig:overview}).
We define perceptual loss as the $L_1$ loss between the maximum MOS (i.e., 5) and the predicted MOS so that minimizing the perceptual loss is equivalent to maximizing the predicted MOS.
The equation for the perceptual loss, $L_\text{per}$, is as follows:
\begin{equation}\label{eq:perceptual_loss}
L_\text{per} = | 5 - \text{predicted MOS} |\,.
\end{equation} 
Then we combine the perceptual loss ($L_\text{per}$) with the conventional loss function ($L_\text{con}$) of a TTS model.
We use Transformer and FastSpeech TTS models to validate our method for the limited training data scenario and knowledge distillation scenario, respectively.

Conventionally, the $L_1$ or $L_2$ distance between the target and the predicted Mel-spectrogram is used as the main loss function for recent TTS.
Based on this loss, denoted as ``Mel loss,'' various loss functions can be used for TTS.  
Transformer TTS \cite{TransformerTTS}, a state-of-the-art TTS model, has a post-net that refines the generated Mel-spectrogram. 
It also has a stop linear layer that predicts the probability of the ``positive stop token,'' which determines the end of the utterance at inference time.
Therefore, the loss function of Transformer TTS consists of $L_2$ Mel losses from before and after the post-net and the binary cross-entropy loss for stop token prediction.
According to \cite{espnet}, the loss function can also include guided attention loss \cite{guided_attn} that forces attention alignment to be diagonal. We also utilize this loss in this paper. 
Finally, we define the conventional loss function $L_\text{con}$ for Transformer TTS as follows:
\begin{equation}\label{eq:transformer_loss}
L_\text{con} = L_\text{bp} + L_\text{ap} + L_\text{stop} + L_\text{attn}\,,
\end{equation} 
where $L_\text{bp}$, $L_\text{ap}$, $L_\text{stop}$, and $L_\text{attn}$ represent the $L_2$ Mel loss from before the post-net, $L_2$ Mel loss from after the post-net, binary cross-entropy loss for stop token prediction, and guided attention loss, respectively.

In this work, FastSpeech \cite{FastSpeech} uses the knowledge distillation to consider the Mel-spectrogram and character duration extracted from the teacher model, pre-trained Transformer TTS, as targets for training.
Accordingly, the loss function of FastSpeech consists of $L_2$ Mel losses from before and after the post-net and the cross-entropy loss for duration prediction.
Finally, the conventional loss function $L_\text{con}$ for FastSpeech is defined as follows:
\begin{equation}\label{eq:fastspeech_loss}
L_\text{con} = L_\text{bp} + L_\text{ap} + L_\text{dur}\,,
\end{equation} 
where $L_\text{dur}$ refers to the cross-entropy loss for duration prediction.

The proposed perceptual loss can be combined with any conventional loss function for TTS, but one issue must be considered when combining the two losses. 
Since the purpose of a MOS test is to evaluate a fully trained speech generation system, the MOS prediction model is trained using Mel-spectrograms from such a system as an input.
Therefore, the Mel-spectrograms predicted in the early stages of TTS training (i.e., from a system that is not fully trained) are out-of-domain data for MOS prediction.
It makes the MOS prediction model unable to predict reliable MOSs for TTS outputs during the early stages of training. 
To address this problem, inspired by \cite{large_AS}, we first set the weight for perceptual loss to a low value and gradually increase it to some extent as the training epoch increases. 
We define the final loss function $L$ as the weighted sum of $L_\text{con}$ and $L_\text{per}$, which is formulated as follows:
\begin{equation}\label{eq:total_loss}
L = \frac{\lambda L_\text{con} + L_\text{per}}{\lambda+1}\,,
\end{equation} 
where $\lambda$, the weight for conventional loss, is initially set to a high value, $\lambda_\text{max}$, and is gradually reduced with a step size of $\gamma$ to a certain level, $\lambda_\text{min}$.
Then, $\lambda$ is defined as follows:
\begin{equation}\label{eq:lambda}
\lambda = \max(\lambda_\text{max} - \gamma \cdot \text{epoch}, \lambda_\text{min})\,,
\end{equation}
where $\lambda_\text{max}$, $\lambda_\text{min}$, and $\gamma$ are determined experimentally.
The parameters of the TTS model are updated to minimize the final loss function.
Under the supervision of perceptual loss, the TTS model can learn to maximize speech quality directly. 

\begin{table}[t]
\caption{A summary of two scenarios for experiments}
\vspace{-0.2cm}
\label{tab:scenario}
\begin{small}
\begin{center}
\renewcommand{\arraystretch}{1.1}
\setlength\tabcolsep{4.5pt}
\begin{tabular}{c|c|c|c}
\hline
No. & Description            & Training data size  & Model  \\ \hline\hline
1   & Limited training data  & Small (6 hours)  & Transformer \\ \hline
2   & Knowledge distillation & Large (18 hours) & FastSpeech  \\ \hline
\end{tabular}
\end{center}
\end{small}
\end{table}

\begin{table*}[t]
\caption{Instructions for the intelligibility test}
\vspace{-0.2cm}
\label{tab:instruction}
\begin{small}
\begin{center}
\renewcommand{\tabcolsep}{0.68mm}
\renewcommand{\arraystretch}{1.1}
\begin{tabular}{c|l}
\hline
Score  & \multicolumn{1}{c}{Instructions}\\\hline\hline
 5 & There is no degradation, and the sentence sounds very clear.\\\hline
\multicolumn{1}{c|}{\multirow{2}{*}{4}} & 
There is degradation in less than or equal to 1/3 of the sentence,\\
& but the meaning of the original sentence is fully conveyed.\\\hline
\multicolumn{1}{c|}{\multirow{2}{*}{3}} & 
There is degradation in less than or equal to 1/3 of the sentence,\\
& but the sentence makes sense itself.\\\hline
\multicolumn{1}{c|}{\multirow{2}{*}{2}} & 
There is degradation in less than or equal to 1/3 of the sentence, and\\
& neither the meaning of the original sentence is fully conveyed nor the sentence makes sense itself.
\\\hline
1 & There is degradation in more than 1/3 of the sentence.\\\hline
\multicolumn{2}{l}{* Degradation: inaccurate, repeated, or skipped pronunciation}\\\hline 
\end{tabular}
\end{center}
\end{small}
\end{table*}

\section{Experiments}
\label{sec:experiments}

\subsection{Dataset}
In experiments, we demonstrate that our method improves the quality of synthesized speech by considering two different scenarios.
A summary of these two scenarios is presented in Table \ref{tab:scenario}.
The first scenario is when limited training data results in speech quality degradation.
For this, we use a 6-hour-long subset of our Korean speech dataset, called the ``Small TTS dataset.''
It consists of 4800 utterances spoken by a professional male speaker in 16-bit PCM WAV format with a sampling rate of 22.05 kHz. 
Further details are available at \href{https://github.com/emotiontts/emotiontts\_open\_db}{https://github.com/emotiontts/emotiontts\_open\_db}.
The dataset is split into 4750, 25, and 25 utterances for training, validation, and testing, respectively.
As discussed in the introduction, oversimplified or inaccurate data generated by a teacher model causes information loss of the target data for the student model, thus degrading the speech quality. This is the second scenario that we consider.
For this, we first train Transformer TTS with another subset of our Korean dataset. 
The subset, called the ``Large TTS dataset,'' contains 13000 utterances recorded by a professional female speaker, corresponding to 18 hours.
We exclude extremely long 48 utterances among them and split the rest into 12822, 65, and 65 utterances for training, validation, and testing, respectively. 
Subsequently, we train FastSpeech using Transformer TTS as the teacher model.

To train the MOS prediction model, MTL-MOSNet, we use the evaluation results of the Voice Conversion Challenge (VCC) 2018 \cite{VCC2018}.
A total of 36 voice conversion systems were submitted to the VCC 2018 and evaluated along with human speech. A total of 20580 utterances were rated by 267 listeners on a scale from 1 (completely unnatural) to 5 (completely natural). 
A ground truth MOS was defined as the average of all the scores an utterance received, and there are 20580 <audio, ground truth MOS> pairs in the evaluation results.
For more details about the dataset, please refer to our previous work \cite{Choi2}.
For each scenario, we train the MOS prediction model using the augmented dataset consisting of both the evaluation results of the VCC 2018 and the TTS dataset as explained in Section \ref{sec:mos_prediction}.

\subsection{Implementation details}
We implement the MOS prediction model using PyTorch and train it on a GTX 1080 Ti GPU. 
The weights for the utterance-level MOS prediction loss, frame-level MOS prediction loss, STC loss, and SD loss are 1, 0.8, 1, and 1, respectively, similar to those in our previous work \cite{Choi}. 
We train the model with the Adam optimizer with a learning rate of $10^{-4}$.
We implement TTS models based on ESPnet \cite{espnet}, which is a widely used end-to-end speech processing toolkit. 
Parallel WaveGAN \cite{PWG} is trained on the same TTS dataset and used as the neural vocoder for each scenario.
Generated audio samples and information about the Korean language are available online at \href{https://wkadldppdy.github.io/perceptualTTS/index.html}{https://wkadldppdy.github.io/perceptualTTS/index.html}.

For the first scenario, we train a Transformer TTS model on the Small TTS dataset using two GTX 1080 Ti GPUs.
Compared to the original paper \cite{TransformerTTS}, we reduce the number of layers from six to three due to the limited training data and adopt character embeddings instead of phoneme ones.
Also, as in \cite{MultiSpeech}, layer normalization is applied to character embeddings.

In the second scenario, we require a teacher model for knowledge distillation. We first train Transformer TTS on the Large TTS dataset using two GTX 1080 Ti GPUs and call it Transformer-L.
Then, we train FastSpeech on a single GTX 1080 Ti GPU using Transformer-L as the teacher model and call it FastSpeech-L. When FastSpeech is perceptually guided, we call it P-FastSpeech-L.

As discussed in Section \ref{sec:perceptually_guided_TTS}, we initially set the value of $\lambda$ to a high value and gradually reduce it to a certain level. 
$\lambda_\text{max}$, $\lambda_\text{min}$, and $\gamma$ in Eq. \ref{eq:lambda} are set to 90, 20, and 1, respectively, for the first scenario and 60, 56, and 0.2, respectively, for the second scenario.

\subsection{Evaluation}
For subjective evaluation, we conduct both naturalness and intelligibility MOS tests, which were also performed in the Blizzard Challenge 2020 \cite{Blizzard2020}.
For each TTS model, 20 listeners rate 25 audio samples, which results in 500 evaluated data points.
For the naturalness test, listeners score each sample in the range from 1 to 5 in increments of 0.5. 
Then we report the MOS of a model as the average of the 500 evaluated data points with a 95\% confidence interval.

In the case of the intelligibility test, the same listeners score each sample on a scale of 1 to 5 in increments of 1. They are provided with input texts of speech samples, unlike in the naturalness test. 
Since we conduct the naturalness test before the intelligibility test, the listeners listen to the samples first without input texts and thus the input texts do not affect the naturalness test.
Unlike general MOS tests, we provide the listeners with input texts for the following reasons.
The speech generated by neural TTS models often suffers from repeated or skipped pronunciations due to imperfect alignment between the text and Mel-spectrogram.
Since the purpose of TTS is to generate speech that exactly matches the input text, scores should be deducted for utterances with such problems.
However, some utterances still make sense even with repeated or skipped words, and they can be problematic when an input text is not available to listeners.
The listeners might misunderstand that the word was repeated for emphasis or hesitation or that the skipped word never existed in the input text, thus giving high scores to the utterance.

Meanwhile, intelligibility is related to aspects such as pronunciation and articulation \cite{assessMOS}.
An in-depth analysis of listeners' scores requires listeners to follow granular evaluation instructions considering those aspects.
By introducing the concepts of adequacy and comprehensibility used in the machine translation field \cite{MT}, we can create more granular instructions.
For TTS, adequacy refers to how well the meaning of the input text is conveyed to the output speech, and comprehensibility refers to how much the output speech is understandable without access to the input text.
Even if the TTS model synthesizes the utterance not exactly according to the text input, the utterance might be adequate and comprehensible. For example, in English, when the script ``I am going to school'' is pronounced as ``I'm going to school,'' the meaning of the input text is still fully conveyed. 
In this case, it is not desirable to assign a low score.
Furthermore, even if the original meaning is not fully conveyed, it is necessary to distinguish between comprehensible and incomprehensible utterances.
Considering the above discussion, we create instructions for the intelligibility test, shown in Table \ref{tab:instruction}.

For objective assessment, we compute the phone error rate (PER) of 200 samples using a Gaussian Mixture Model-Hidden Markov Model (GMM-HMM) phone recognizer. 
We use the Kaldi toolkit to train the phone recognizer on the combination of both the Small and Large TTS datasets.
The 200 sentences consist of 47 ``long'' sentences (excluded from the Large TTS dataset) and 153 relatively ``short'' sentences (60 sentences from the Large TTS test data and 93 sentences not in any TTS dataset).
Here, ``long'' sentences have 188 phonemes, whereas ``short'' sentences have 43 phonemes on average.
Since the length of the input sentence can affect the quality of synthesized speech, we report both PERs for long and short sentences separately in addition to the overall PER.

\begin{figure*}[t]
    \centering
    \begin{small}
    \renewcommand{\tabcolsep}{0.8mm}
    \renewcommand{\arraystretch}{1.0}
        \begin{tabular}{ccc}
            \includegraphics[trim=0.4cm 0.3cm 0.1cm 0.4cm, clip=true, width=0.315\textwidth]{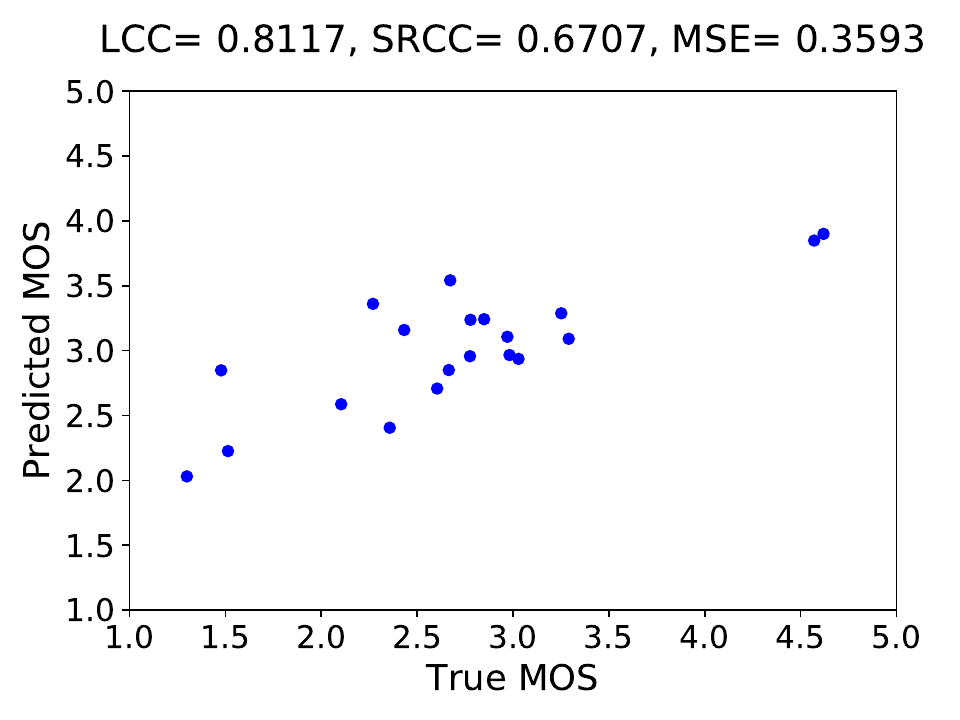} &
            \includegraphics[trim=0.4cm 0.3cm 0.1cm 0.4cm, clip=true, width=0.315\textwidth]{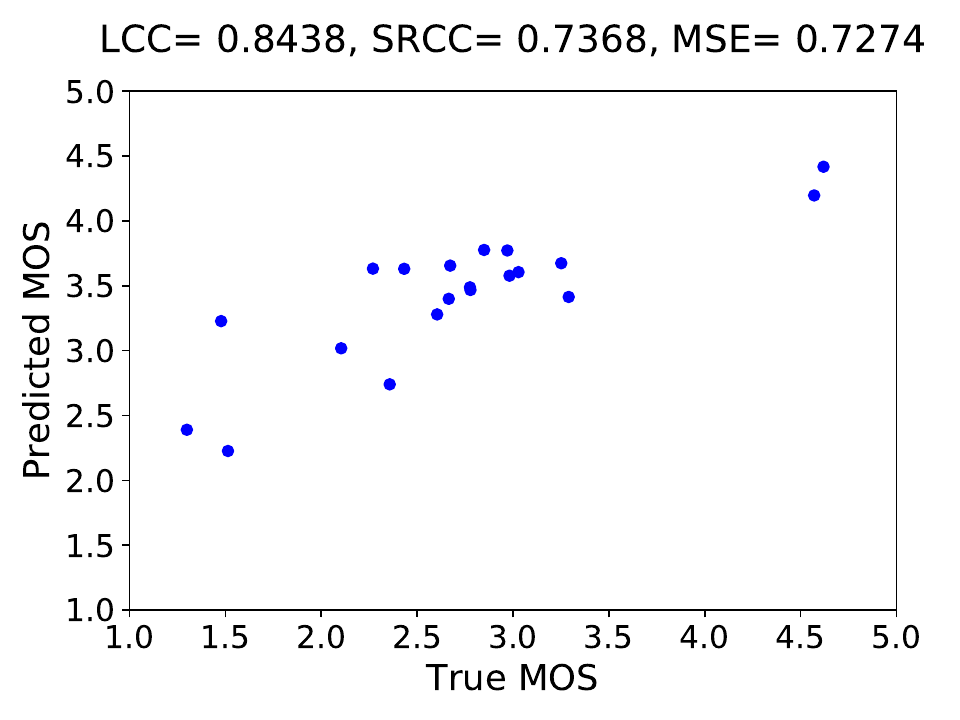} &
            \includegraphics[trim=0.05cm 0.3cm 0.45cm 0.4cm, clip=true, width=0.315\textwidth]{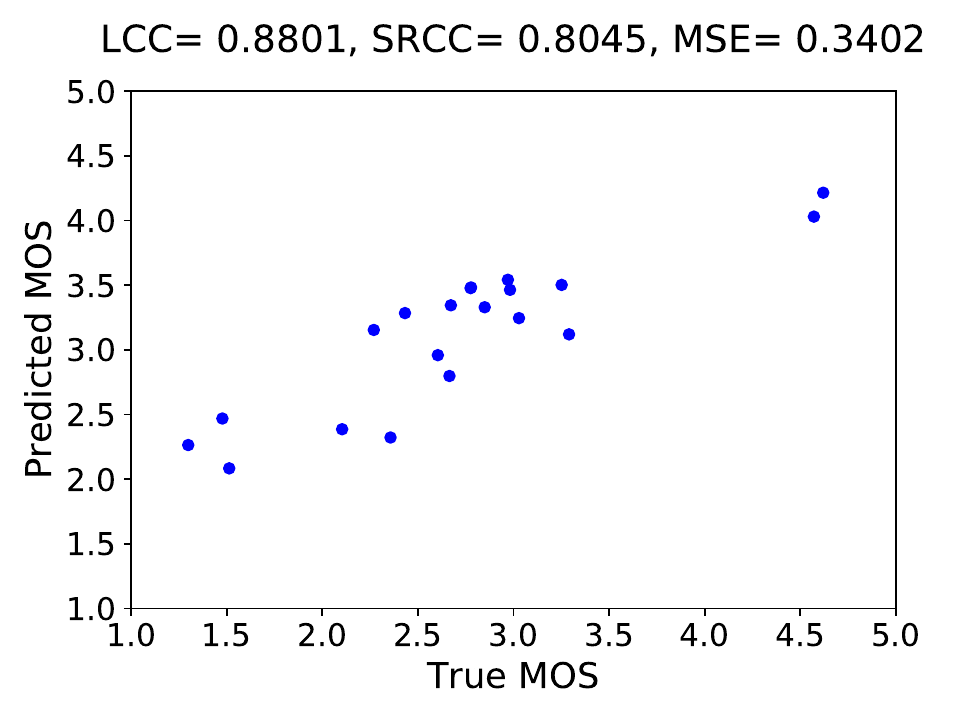} \\
            (a) w/o data augmentation & (b) w/ data augmentation (Small) & (c) w/ data augmentation (Large)
        \end{tabular}
    \end{small}
\caption{Scatter plots of MTL-MOSNet trained on the dataset (a) without augmentation, (b) with augmentation using the Small TTS dataset, and (c) with augmentation using the Large TTS dataset.}
\label{fig:MOS_prediction}
\end{figure*}

\section{Results and discussion}
This section presents the results and analysis for two scenarios: limited training data and knowledge distillation.
Note that the findings of this study have to be seen in the light of some limitations. First, as in the majority of multi-task learning studies, the loss weights are determined heuristically. This not only makes the experiments time-consuming but also can yield suboptimal results. Second, a fundamental domain mismatch remains between the Mel-spectrogram generated by TTS and that of the MOS prediction dataset. As discussed in Section \ref{sec:method}, to handle domain mismatch, we augment the MOS prediction dataset with TTS speech data and initially set the weight for perceptual loss to a low value. However, these methods cannot eliminate the domain mismatch completely.
Finally, during the subjective intelligibility test, there is bias in listening behaviors caused by the prior access to the input texts. 
Manual dictation by human subjects can directly address the problem since scripts are not provided to human subjects.
Instead, we can adopt automatic phone recognition which can serve as manual dictation by human subjects.
Therefore, PER results could compensate for the bias in the subjective intelligibility test results, but the bias itself still remains. 
In the future, we will develop a more advanced approach to overcome these limitations.

\begin{table}[t]
\caption{The naturalness MOSs and PERs on the Small TTS dataset}
\vspace{-0.2cm}
\label{tab:results_maleDB}
\begin{small}
\begin{center}
\renewcommand{\tabcolsep}{1.2mm}
\renewcommand{\arraystretch}{1.1}
\begin{tabular}{c|c|ccc}
\hline
\multirow{2}{*}{System} & \multirow{2}{*}{MOS} & \multicolumn{3}{c}{PER (\%)}\tabularnewline\cline{3-5}
 & & long & short & overall \\\hline\hline
Transformer-S & 2.71 $\pm$ 0.091 & 62.03 & 12.66 & 40.88 \\\hline
P-Transformer-S & 3.75 $\pm$ 0.071 & 47.77 & 6.46 & 30.08\\\hline
\end{tabular}
\end{center}
\end{small}
\end{table}

\subsection{Results for MOS prediction}
\label{sec:MOS_prediction_results}

Before discussing the TTS results, we present the MOS prediction results. For the limited training data scenario, we train MTL-MOSNet on the augmented dataset containing both the evaluation results of VCC 2018 and the Small TTS dataset. For the knowledge distillation scenario, MTL-MOSNet is trained on the dataset augmented with the Large TTS dataset. Note that the final goal of these MOS prediction models is to improve the performance of a TTS model by providing perceptual loss. However, the goal can be achieved only if the MOS prediction models work properly. Therefore, we focus on validating that the MOS prediction models predict reasonable MOSs, including MTL-MOSNet trained without the augmented dataset.

To specifically validate the generalization ability of MOS prediction models, we test the models on MOS evaluation results from the VCC 2016 \cite{VCC2016}. There is no specification of the utterances or raters for those MOS evaluation results, which means that we can only measure system-level performance. The performance of MOS prediction is measured in terms of the linear correlation coefficient (LCC) \cite{Pearson}, Spearman's rank correlation coefficient (SRCC) \cite{Spearman}, and mean squared error (MSE). Fig. \ref{fig:MOS_prediction} shows scatter plots of (a) MTL-MOSNet trained on the dataset without augmentation, (b) MTL-MOSNet trained on the dataset augmented with Small TTS dataset, and (c) MTL-MOSNet trained on the dataset augmented with Large TTS dataset. These results show that pre-trained MOS prediction models predict reasonable MOSs from input Mel-spectrograms (an LCC of $>$ 0.8 generally suggests a strong positive association between two variables).

\begin{figure}
\centerline{\includegraphics[trim=0.2cm 6.8cm 0.8cm 0.8cm, clip=true,width=\columnwidth]{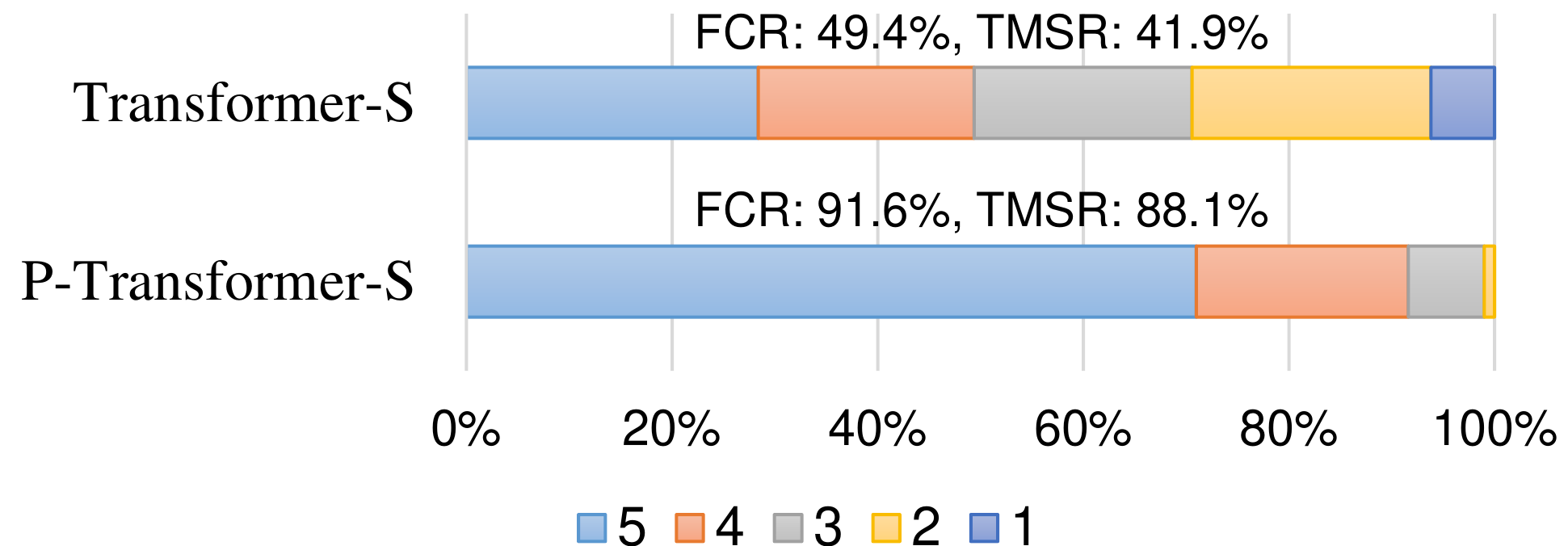}}
\caption{Intelligibility test results on the Small TTS dataset.} 
\label{fig:bar_maleDB}
\end{figure}

\subsection{Results for limited training data}
\label{sec:maleDB}

We compare the performance of Transformer TTS and perceptually guided Transformer TTS trained on the Small TTS dataset. We denote these models as Transformer-S and P-Transformer-S, respectively. Table \ref{tab:results_maleDB} lists the naturalness MOSs and PERs of them.
We can see that our method, P-Transformer-S, increases the naturalness MOS by more than 1.
The p-value of the paired t-test is lower than 0.01, indicating that the improvement is quite significant (a p-value of $<$ 0.05 is taken as statistically significant).
The PER decreases for both long and short sentences, and we obtain a relative improvement of 26.4\% overall. 
Here, both PERs for long sentences are high (i.e., over 40\%), which is not only because Transformer TTS generates unstable attention alignments when converting long sentences into speech \cite{TTSattn} but also because it lacks training data.

Fig. \ref{fig:bar_maleDB} shows the subjective intelligibility test results on the Small TTS dataset. It clearly demonstrates that P-Transformer-S outperforms Transformer-S. 
For in-depth analysis using the instructions in Table \ref{tab:instruction}, we define the ratio of the ``fully conveyed'' as $FCR = (N_4+N_5)/N_\text{tot}$ and the ratio of the ``though makes sense'' as $TMSR = N_3/(N_1+N_2+N_3)$. 
Here, $N_n$ is the number of evaluated data points with scores of $n$, and $N_\text{tot}$ is the total number of evaluated data points (i.e., 500).
When the denominator of $TMSR$ is less than 3, we do not define $TMSR$ and denote it as ``-'' because the sum of N1, N2, and N3 is too small.
$FCR$ focuses on highly intelligible evaluated data where the meaning of the original sentence is fully conveyed, whereas $TMSR$ focuses on evaluated data where the sentence at least makes sense even though the meaning of the original sentence is not fully conveyed. 
Therefore, we can say that higher $FCR$ represents better adequacy and higher $TMSR$ represents better comprehensibility. 
Since $FCR$ increases from 49.4\% to 91.6\% and $TMSR$ increases from 41.9\% to 88.1\%, we can say that our method improves the intelligibility of Transformer-S.

\begin{table}[t]
\caption{The naturalness MOSs and PERs on the Large TTS dataset}
\vspace{-0.2cm}
\label{tab:results_femaleDB}
\begin{small}
\begin{center}
\renewcommand{\tabcolsep}{1.2mm}
\renewcommand{\arraystretch}{1.1}
\begin{tabular}{c|c|ccc}
\hline
\multirow{2}{*}{System} & \multirow{2}{*}{MOS} & \multicolumn{3}{c}{PER (\%)}\tabularnewline\cline{3-5}
 & & long & short & overall \\\hline\hline
GT (Mel) & 4.09 $\pm$ 0.066 & 9.47 & - & -\\\hline
Transformer-L & 4.06 $\pm$ 0.072 & 22.29 & 3.98 & 14.45\\\hline
P-Transformer-L & 4.01 $\pm$ 0.070 & 22.54 & 3.75 & 14.49 \\\hline
FastSpeech-L & 3.40 $\pm$ 0.085 & 12.50 & 4.24 & 8.96 \\\hline
P-FastSpeech-L & 3.78 $\pm$ 0.077 & 11.50 & 4.05 & 8.31 \\\hline
\end{tabular}
\end{center}
\end{small}
\end{table}

\subsection{Results for knowledge distillation}
\label{sec:femaleDB}

The second, fourth, and fifth rows of Table \ref{tab:results_femaleDB} present the naturalness MOSs and PERs of Transformer-L, FastSpeech-L, and P-FastSpeech-L, respectively.
In terms of naturalness MOS, P-FastSpeech-L outperforms FastSpeech-L with a gap of 0.38 (p-value < 0.01), which gets closer to the teacher model (Transformer-L).
P-FastSpeech-L achieves a 7.25\% relative improvement in the overall PER. 
Here, as opposed to the PERs on short sentences, the PERs on the long sentences are almost half that of Transformer-L because FastSpeech-L is more robust to the length of input text.
As noted in numerous studies such as \cite{TTSattn}, the attention mechanism of a neural TTS model often fails to align between the input text and output Mel-spectrogram in the latter part when the text is long. Unlike Transformer-L, FastSpeech-L does not use an attention mechanism. Therefore, it can produce speech more reliably than Transformer-L even when the input text is long.

\begin{figure}
\centerline{\includegraphics[trim=0.3cm 0.8cm 3.0cm 0.7cm, clip=true,width=\columnwidth]{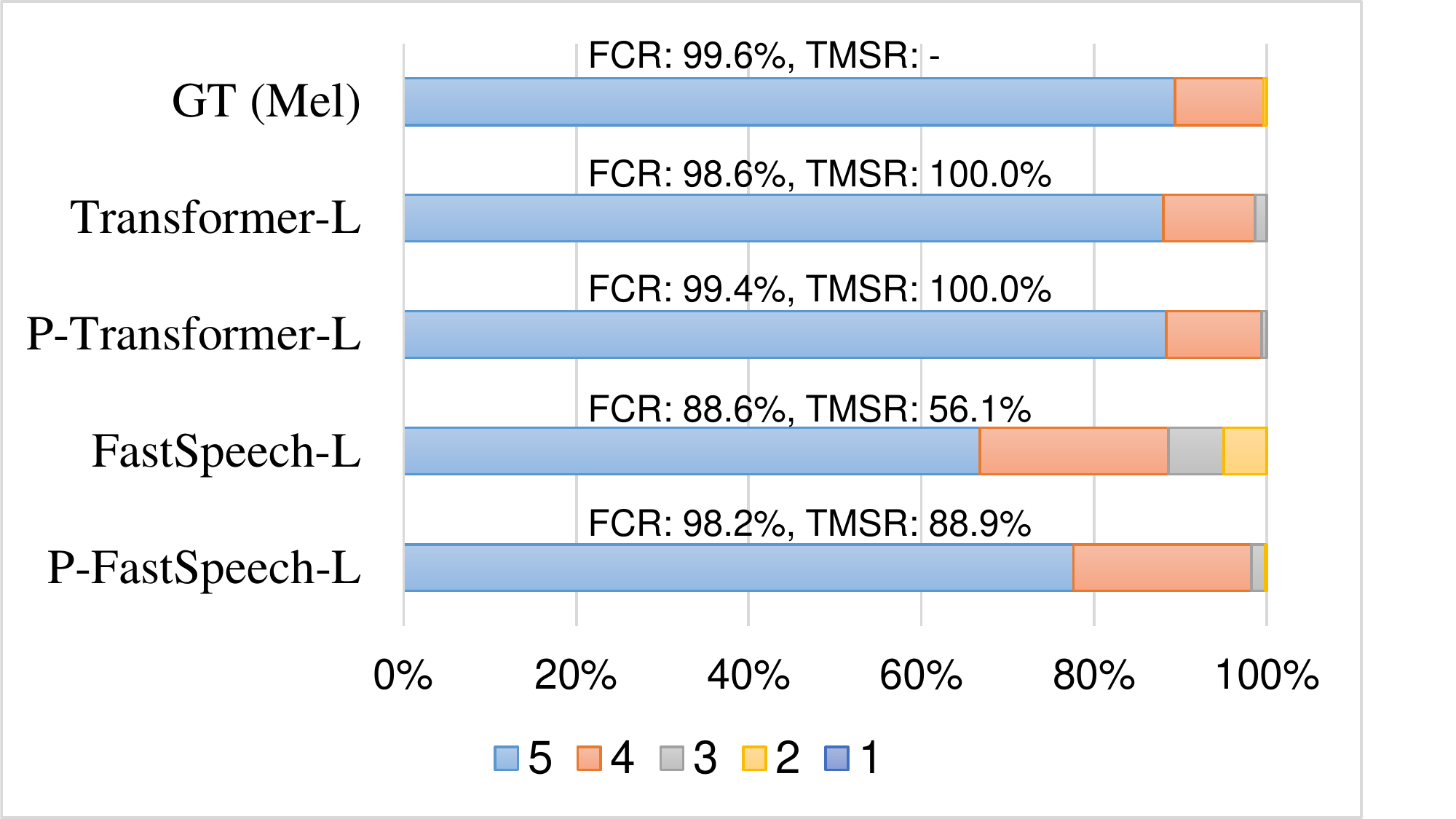}}
\caption{Intelligibility test results on the Large TTS dataset.} 
\label{fig:bar_femaleDB}
\end{figure}

Fig. \ref{fig:bar_femaleDB} shows the intelligibility test results on the Large TTS dataset. 
In that $FCR$ increases from 88.6\% to 98.2\% and $TMSR$ increases from 56.1\% to 88.9\%, P-FastSpeech-L outperforms FastSpeech-L. 
It is comparable to Transformer-L, which shows $FCR$ of 98.6\% and $TMSR$ of 100.0\%.

Besides the second scenario, we perform an additional experiment to investigate whether the proposed method is also effective for the state-of-the-art TTS model, Transformer TTS.
We train perceptually guided Transformer TTS and call it P-Transformer-L. After that, we compare P-Transformer-L with Transformer-L and the system called ``GT (Mel).''
In GT (Mel), we convert the ground truth Mel-spectrogram into a waveform using Parallel WaveGAN.

The results of GT (Mel), Transformer-L, and P-Transformer-L are presented in the first, second, and third rows of Table \ref{tab:results_femaleDB}, respectively. 
In terms of naturalness MOS, p-values for all possible pairs between the three systems are higher than 0.29.
Therefore, the pairwise differences between naturalness MOSs of the three systems are not statistically significant, and we cannot tell which one is better. 
More specifically, although P-Transformer-L shows a lower MOS than Transformer-L, we cannot conclude that the proposed method degrades the naturalness of Transformer-L.
Because the proposed method works effectively in the case of Transformer-S and FastSpeech-L, we analyze why it cannot improve Transformer-L by focusing on the differences from the other two models.
In contrast to Transformer-S, Transformer-L is trained with enough data.
This difference results in better naturalness and intelligibility compared to Transformer-S. Then, for especially long sentences, whereas the synthesized speech of Transformer-S shows poor naturalness and intelligibility from the beginning to the end of a sentence, the synthesized speech of Transformer-L shows high naturalness and intelligibility at least before the latter part of a sentence. 
As discussed earlier, the latter part of the synthesized speech can be unintelligible since Transformer-L uses an attention mechanism. 
However, FastSpeech-L, which does not use an attention mechanism can generate speech that shows high intelligibility until the end of a long input text.

P-Transformer-L does not outperform Transformer-L because of this characteristic of Transformer-L. 
For long text input, Transformer-L often generates speech that has high quality up to a certain point but is unintelligible at the end of the sequence. 
In the case of such speech, the quality varies significantly between the former and latter parts. 
Then, a single MOS value alone is insufficient to evaluate the whole sentence, and thus, the perceptual loss based on a single MOS loses its power to guide the TTS model.

In terms of the PER, P-Transformer-L shows a relative degradation of 1.12\% for long sentences but achieves a relative improvement of 5.78\% for short sentences. 
The PER of the GT (Mel) system is only reported for long sentences since there are no recordings for 93 short sentences.
The intelligibility test results are shown in the first three rows of Fig. \ref{fig:bar_femaleDB}. 
By perceptual training, $FCR$ of Transformer-L increases from 98.6\% to 99.4\%, which is only 0.2\% lower than that of GT (Mel). 
These results show that our method improves the intelligibility of Transformer-L for short sentences. 

\subsection{Ablation studies}
We conduct ablation studies to verify the effectiveness of the proposed data augmentation method for MOS prediction. 
We train both Transformer-S and FastSpeech-L under the supervision of MTL-MOSNet trained only on the evaluation results of VCC 2018 (i.e., without the TTS dataset). 
The results are in Table \ref{tab:ablation_studies}. 
For the limited training data scenario, the overall PER for P-Transformer-S without data augmentation is 35.39\%, which is worse than 30.08\% for P-Transformer-S but better than 40.88\% for Transformer-S.
For the knowledge distillation scenario, the overall PER for P-FastSpeech-L without data augmentation is 8.56\%, which is worse than 8.31\% for P-FastSpeech-L but better than 8.96\% for FastSpeech-L.
As can be seen from the results, the proposed method can reduce the overall PER even without data augmentation. Moreover, we can observe that using data augmentation leads to a larger relative improvement in PER.

\begin{table}[t]
\caption{Results for ablation studies}
\vspace{-0.2cm}
\label{tab:ablation_studies}
\begin{small}
\begin{center}
\renewcommand{\tabcolsep}{0.9mm}
\renewcommand{\arraystretch}{1.2}
\begin{tabular}{c|c|c}
\hline
TTS Scenario & System  & Overall PER (\%)\\\hline\hline
 \multicolumn{1}{c|}{\multirow{4}{*}{Limited training data}} & Transformer-S & 40.88\\\cline{2-3}
 & P-Transformer-S w/o & \multicolumn{1}{c}{\multirow{2}{*}{35.39}}\\
& data augmentation &\\\cline{2-3}
& \textbf{P-Transformer-S} & \textbf{30.08}\\\hline\hline
 \multicolumn{1}{c|}{\multirow{4}{*}{Knowledge distillation}} & FastSpeech-L & 8.96\\\cline{2-3}
 & P-FastSpeech-L w/o & \multicolumn{1}{c}{\multirow{2}{*}{8.56}}\\
& data augmentation &\\\cline{2-3}
& \textbf{P-FastSpeech-L} & \textbf{8.31}\\ \hline
\end{tabular}
\end{center}
\end{small}
\end{table}

\section{Conclusion}

We proposed a perceptual training method for a TTS model to improve the speech quality independently and efficiently.
We first trained the MOS prediction model on the augmented data and then used the model to calculate the perceptual loss for the TTS model.
Under the supervision of perceptual loss, the TTS model learned to maximize the perceptual speech quality directly.
The experimental results for two scenarios show that the proposed method improves the previous TTS models in terms of naturalness and intelligibility.
In future work, we will develop a sophisticated approach to automatically find the optimal loss weights instead of simply using heuristically determined values.
We will also extend our study to other speech generation tasks, such as multi-speaker TTS.

\begin{IEEEbiography}[{\includegraphics[width=1in,height=1.25in,clip,keepaspectratio]{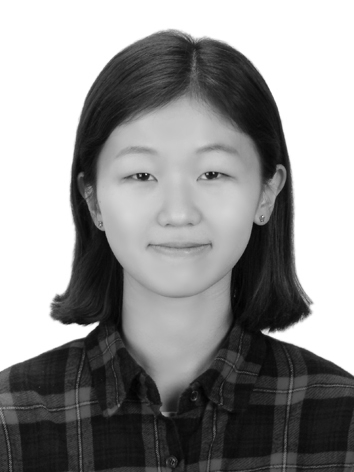}}]{YEUNJU CHOI} received the B.S. and M.S. degrees in electrical engineering from Korea Advanced Institute of Science and Technology (KAIST), Daejeon, South Korea, in 2016 and 2018, respectively, where she is currently pursuing the Ph.D. degree. Her research interests include deep learning and signal processing for speech synthesis, voice conversion, speaker verification, spoofing detection, and speech quality prediction.
\end{IEEEbiography}

\begin{IEEEbiography}[{\includegraphics[width=1in,height=1.25in,clip,keepaspectratio]{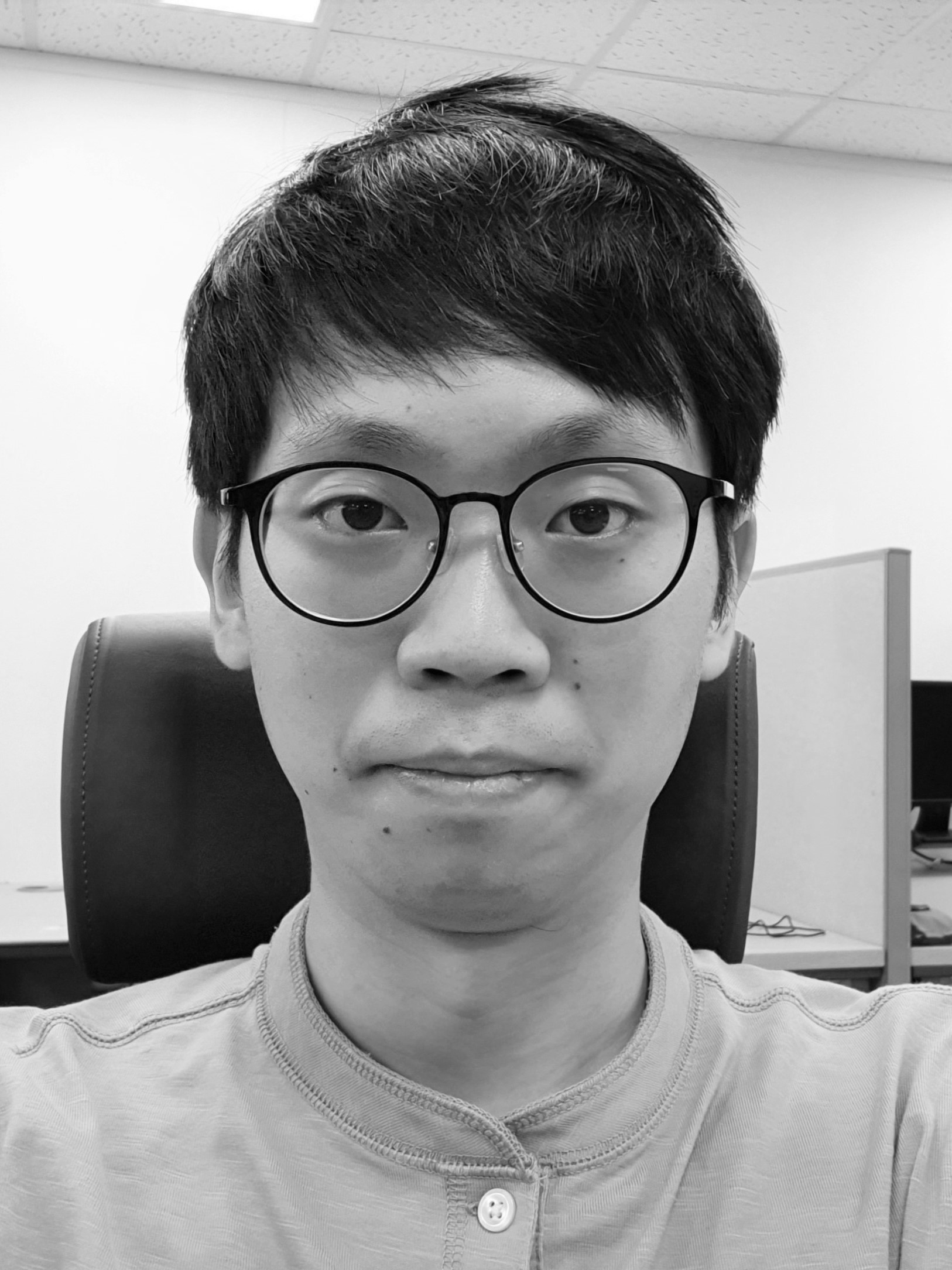}}]{YOUNGMOON JUNG} received the B.S. degree in electronic engineering from Sogang University, Seoul, South Korea, in 2016, and the Ph.D. degree in electrical engineering from Korea Advanced Institute of Science and Technology (KAIST), Daejeon, South Korea, in 2022. 
His research interests include deep learning and signal processing for speaker recognition, speech synthesis, speech enhancement, and voice activity detection.
\end{IEEEbiography}

\begin{IEEEbiography}[{\includegraphics[width=1in,height=1.25in,clip,keepaspectratio]{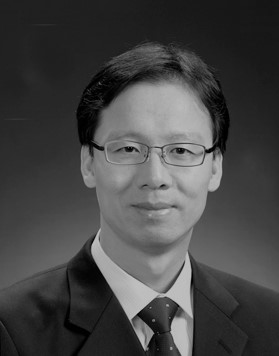}}]{YOUNGJOO SUH} received the B.S. and M.S. degrees in electronics engineering from Kyungpook National University, Daegu, South Korea in 1991 and 1993, respectively and Ph.D. degree in information and communications engineering from Korea Advanced Institute of Science and Technology (KAIST), Daejeon, South Korea in 2006. From 1993 to 1998, he was a Researcher with the Spoken Language Processing Laboratory, Electronics and Telecommunications Research Institute, South Korea. From 2000 to 2002, he was a Co-Founder and Team Manager with CoreVoice Inc., Seoul, South Korea. From 2006 to 2021, he was a Research Associate Professor and then a Contract Researcher with the School of Electrical Engineering, KAIST. Since 2021, he has been a Director with Voice Group, Konan Technology Inc., Seoul, South Korea. His research interests include deep learning-based signal processing for speech recognition, speech synthesis and voice conversion.
\end{IEEEbiography}

\begin{IEEEbiography}[{\includegraphics[width=1in,height=1.25in,clip,keepaspectratio]{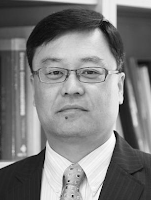}}]{HOIRIN KIM} (M'92) received the B.S. degree in electronics engineering from Hanyang University, Seoul, South Korea, in 1984, and the M.S. and Ph.D. degrees in electrical engineering from Korea Advanced Institute of Science and Technology (KAIST), Daejeon, South Korea, in 1987 and 1992, respectively.

From 1987 to 1999, he was a Senior Researcher with the Spoken Language Processing Laboratory, Electronics and Telecommunications Research Institute, South Korea. From 1994 to 1995, he was on leave with ATR-ITL, Kyoto, Japan, as a Visiting Researcher. From 2000 to 2009, he was an Associative Professor with the Information and Communications University, South Korea. From 2006 to 2007, he was on leave with the INC, University of California at San Diego, San Diego, USA, as a Visiting Scholar. Since 2009, he has been a Professor with the School of Electrical Engineering, KAIST.
His research interests include signal processing for speech and speaker recognition, speech synthesis, audio indexing and retrieval, and spoken language processing.
\end{IEEEbiography}

\EOD

\end{document}